\documentclass[twocolumn,showpacs,preprintnumbers,amsmath,amssymb,prl,superscriptaddress]{revtex4-1}
\usepackage{amsfonts}
\usepackage[english]{babel}
\usepackage[T1]{fontenc}
\usepackage{times}
\usepackage{mathrsfs}
\usepackage{graphicx}% Include figure files
\usepackage{dcolumn}% Align table columns on decimal point
\usepackage{bm}% bold math
\usepackage[colorlinks,bookmarks=true,citecolor=blue,linkcolor=red,urlcolor=blue]{hyperref}
\usepackage[tight, FIGTOPCAP, hang, raggedright, nooneline]{subfigure}
\usepackage{hyperref}

\begin{document}
\title{Dynamics of a two-dimensional quantum spin liquid:\\ signatures of emergent Majorana fermions and fluxes}
\author {J.~Knolle}
\affiliation{Max Planck Institute for the Physics of Complex Systems, D-01187 Dresden, Germany} 
\author{D.~L.~Kovrizhin}
\affiliation{Max Planck Institute for the Physics of Complex Systems, D-01187 Dresden, Germany} 
\affiliation{ T.C.M. Group, Cavendish Laboratory, J. J. Thomson Avenue, Cambridge CB3 0HE, United Kingdom, and\\ Imperial College London, London, SW7 2AZ, United Kingdom}
\author{J.~T.~Chalker}
\affiliation{Theoretical Physics, Oxford University, 1, Keble Road, Oxford OX1 3NP, United Kingdom}
\author{R.~Moessner}
\affiliation{Max Planck Institute for the Physics of Complex Systems, D-01187 Dresden, Germany}

%\begin{abstract}

%\end{abstract}

\date{\today}

\maketitle

\textbf{
Topological states of matter present a wide variety of striking new phenomena. Prominent  among these is the fractionalisation of electrons into unusual particles: Majorana fermions \cite{Kitaev2006}, Laughlin quasiparticles \cite{Laughlin83} or magnetic monopoles \cite{CMS2008}. Their detection, however, is fundamentally complicated by the lack of any local order, such as, for example, the magnetisation in a ferromagnet. While there are now several instances of candidate topological spin liquids \cite{Lee2008}, %with possible fractionalised states, 
%there is a woeful shortage of diagnostics for their identification 
their identification remains challenging \cite{kagome}. Here, we provide a complete and exact theoretical study of the dynamical structure factor of a two-dimensional quantum spin liquid in gapless and gapped phases. We show that there are direct signatures -- qualitative and quantitative -- of the Majorana fermions and gauge fluxes emerging in Kitaev's honeycomb model. These include counterintuitive manifestations of quantum number fractionalisation, such as a neutron scattering response with a gap even in the presence of gapless excitations, and a sharp component despite the fractionalisation of electron spin. Our analysis identifies new varieties of the venerable X-ray edge problem and explores connections to the physics of quantum quenches.}

The study of spin liquids has been central to advancing our understanding of correlated phases of quantum matter ever since Anderson's proposal of the resonating valence bond (RVB) liquid state \cite{Anderson1973}, which provided, via the detour of high-temperature superconductivity, an early instance of a fractionalised topological state \cite{RokKiv88,RM_SLS2001}. 
%There the electrons break up into spinons and holons -- excitations carrying separately their spin and charge quantum numbers. 
More recent manifestations hold the promise of realising an architecture of quantum computing robust against decoherence \cite{RMP2008}.

\begin{figure}[tb]
\begin{centering}
\includegraphics[width=0.9\columnwidth]{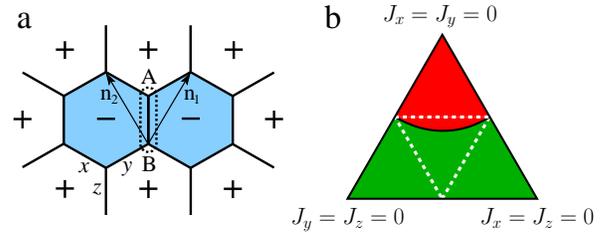}
\par\end{centering}
\caption{
%(Colour online)
{\bf The Kitaev honeycomb model.}
(a) The structure of the model on the honeycomb lattice with two sublattices (labeled $A$ and $B$) and three bond directions (denoted $x,y,z$). The calculation of the dynamical response can be mapped to a local quantum quench, in which two adjoining $Z_2$ fluxes, shown in blue, are inserted. (b) The ground state and dynamical phase diagrams of the model: the phase with gapless fermion excitations fills the central triangle while gapped phases  occupy the three outer triangles. The dynamical response $S^{zz}({\bf q},\omega)$ includes a contribution sharp in $\omega$ in the red region, but not in the green region.
\label{lattice}}
\end{figure}

Investigations of such topological states are hampered by the lack of suitable approaches, with numerical methods limited to small system sizes, to models with a robust excitation gap, or ones that avoid the sign problem in quantum Monte Carlo. A 
benchmark
%valuable counter-point 
is offered by the Kitaev model \cite{Kitaev2006}, which can be used as a representative example of an entire class of quantum spin liquids (QSL). While being a minimal model, it combines a raft of desirable features. First, it is described by a simple Hamiltonian involving only nearest-neighbor interactions on the honeycomb lattice (Fig. \ref{lattice}), by virtue of which it is a promising candidate for realisation in materials physics \cite{Jackeli2009}, or in cold atom implementations of quantum simulators \cite{Duan2003}. Second, it harbours two distinct topological quantum spin liquid phases, with either gapless or gapped Majorana fermion excitations. Finally, its solution can be reduced to the problem of Majorana fermions hopping in the background of an emergent static gauge field. 

This remarkable feature permits, at least in principle, even an analysis of the model's dynamical properties, as noted in a seminal paper by Baskaran and co-workers \cite{Baskaran2007}, who pointed out an unexpected connection to the X-ray edge problem \cite{Gogolin2004}, results of which were used to extract asymptotic correlators of related models \cite{Tikhonov2010,Tikhonov2011}. 
This problem,
%The X-ray edge problem, represents one of the first examples of a quantum quench, where a non-interacting system of electrons evolves in time after an abrupt change of a local external potential. This problem, 
whose {\it tour de force} exact solution was obtained by Nozieres et.al.~\cite{Nozieres1969}, is one of the cornerstones of condensed matter physics, linked to the discovery of Anderson's `orthogonality catastrophe' \cite{Anderson1967}, and a foundation for our understanding of local quantum quenches.
%, and provided the basis for another marked achievement of the last decade --- the theory of non-linear Luttinger liquids \cite{Glazman}. 
%Following Baskaran's proposal the results of the X-ray edge physics were used to extract asymptotic correlations in two related models \cite{Tikhonov2010,Tikhonov2011}. Here, we show that the physics is considerably richer.

The possibility of accessing dynamical properties of spin-liquids is of particular importance as these contain information on fractionalised quasiparticles, and their theoretical study is topical in view of recent neutron scattering investigations of candidate QSL compounds \cite{kagome,Coldea2012,Gegenwart2012}. Indeed, the $S = 1/2$ spinons in the Heisenberg chain were most impressively visualised \cite{ruegg,Tennant} by an analysis of experiments based on the exact Bethe-ansatz solution, specific to one dimension. 
Our work provides this information in complete detail for the first time for a fractionalised quantum spin liquid in more than one dimension. Through the connections to quantum quenches and the physics of Majorana Fermions which appear in our discussion, it cements the central role played by the Kitaev model for our understanding of correlated and topological phases. 

%Dynamical properties are of particular interest as they contain information on fractionalised quasiparticles, and their theoretical study is topical in view of  recent neutron scattering investigations of candidate spin liquid compounds \cite{kagome,Coldea2012,Gegenwart2012}.
%%The study of dynamical properties is timely given the recent neutron scattering activities on candidate spin liquid compounds \cite{kagome,Coldea2012,Gegenwart2012}. They are of exceptional interest as they probe not just the ground state but also involve the spectrum of excited states, which contains precious information on the fractionalised quasiparticles. 
%Indeed, the $S = 1/2$ spinons in the Heisenberg chain were most impressively visualised \cite{ruegg,Tennant} by an analysis of experiments based on the exact Bethe ansatz solution, specific to one dimension.

%Here, following a brief introduction to the Kitaev model (Fig.~\ref{lattice}), we present %our
%numerically exact results for its full dynamic structure factor in both gapless and gapped spin-liquid phases %and in particular discuss the fingerprints of the emergent quasiparticles: gauge fluxes and Majorana %fermions, Fig.~\ref{threepanels}. We explore the connections of our problem to the physics of local quantum %quenches and the $X$-ray edge and close with a discussion of experiment. 

\begin{figure*}[!t]
\begin{centering}
\includegraphics[width=1.95\columnwidth]{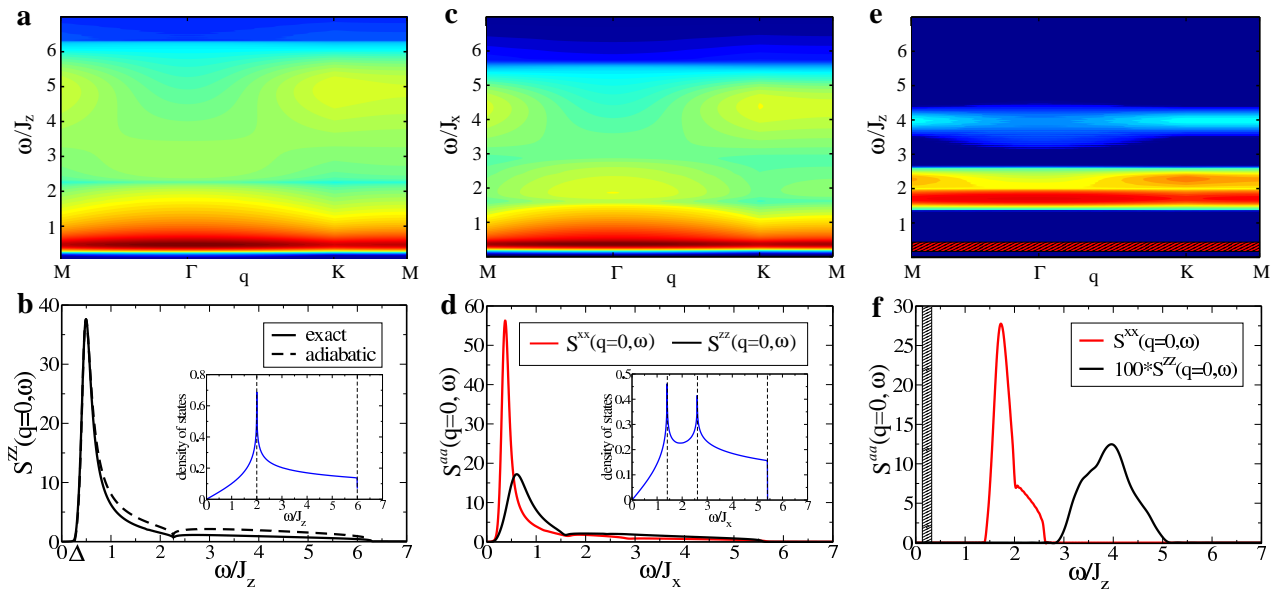}
\par\end{centering}
\caption{
%(Colour online). 
{\bf The exact dynamical structure factor of the Kitaev spin liquids.}
Total dynamical structure factor $S(\mathbf{q},\omega)=\sum_{a}S^{aa}(\mathbf{q},\omega)$ and the inequivalent components $S^{aa}(
%\mathbf{q=0}
0,\omega)$, as would be measured in inelastic neutron scattering and electron spin resonance, respectively, evaluated for three points in the phase diagram: (a+b) the symmetric point ($J_{x}{=}J_{y}{=}J_{z}$); (c+d) a gapless asymmetric point  ($J_{x}{=}J_{y},\, J_z{=}0.7J_x$) and (e+f) a gapped point ($J_{x}{=}J_{y}{=}0.15J_{z}$). Top: 
$S(\mathbf{q},\omega)$ on a logarithmic colour scale as a function of $\omega$ along the cut $\mathrm{M\Gamma K M}$ through the Brillouin zone. Bottom: dynamical susceptibility $
%\chi
S^{aa}(
%\mathbf{q}=
0,\omega)$ for $a=z,x$ at the same values of the exchange. Comparison with the adiabatic response as explained in the main text is given in panel b (black dashed line). The dashed line in (f) indicates the delta-function contribution to the response, present only in the region of the dynamical phase diagram coloured red in Fig. 1b. The insets to panels (b) and (d) show the density of states of the matter fermions.\label{threepanels}}
\end{figure*}

{\it The model.} In the Kitaev model spin-half degrees of freedom at sites $j$ of a honeycomb lattice interact via nearest-neighbour Ising exchange $J_a$. Frustration and quantum fluctuations stem from linking the anisotropy direction $a=x,y,z$ in spin space to the bond direction in real space (Fig.~\ref{lattice}a), a form of spin-orbit coupling. With Pauli matrices $\hat{\sigma}_j^{a}$ and using $\langle ij \rangle_a$ to indicate two sites sharing an $a$ bond, the Hamiltonian is
%The basic degrees of freedom are spins-1/2 residing on vertices $j$ of a honeycomb lattice. The spins, represented by Pauli-matrices $\hat{\sigma}_j^{a}$, $a=x,y,z$, interact via a nearest-neighbour Ising exchange $J_a$ with the anisotropy direction in spin space being linked to that of the bond in real space (``spin-orbit coupling'', Fig.~\ref{lattice}):
%01
\begin{equation}\label{HkitaevSpin}
{H}  =  -J_x\sum_{\langle i j  \rangle_x}\hat{\sigma}_i^x \hat{\sigma}_j^x  -J_y\sum_{\langle i j  \rangle_y}\hat{\sigma}_i^y\hat{\sigma}_j^y-J_z\sum_{\langle i j  \rangle_z}\hat{\sigma}_i^z \hat{\sigma}_j^z. 
\end{equation}
The ground states of Eq.~(\ref{HkitaevSpin}) fall into two classes \cite{Kitaev2006} -- gapped and gapless spin-liquids -- depending on the relative values of the 
$J_a$ (Fig.~\ref{lattice}b). Their emergent independent degrees of freedom are static $\mathbb{Z}_2$ gauge fluxes threading the plaquettes of the honeycomb lattice and Majorana fermions that hop between sites in this gauge field.
%occupying its sites. The dynamics of the latter takes the form of a quadratic hopping problem with the signs of the tunnelings fixed by the flux configuration (modulo gauge transformations).

The model is solved \cite{Kitaev2006} by introducing four Majorana fermions $\hat{c}_{i}, \hat{b}_{i}^{x},\hat{b}_{i}^{y},\hat{b}_{i}^{z}$ at each site
%, $a=x,y,z$ on every lattice site $i$, with commutators of the form $\{\hat{b}_i^{a}, b^{a_2}_j\}=2\delta_{ij}\delta_{a_1 a_2}$ . The spins are represented in terms of $c_{i}$ and $\hat{b}^{a}_i$ via $
and representing spins as
$\hat{\sigma}_i^a=i \hat{c}_i \hat{b}_i^a$. 
%Using $\langle ij \rangle_a$ to indicate two sites sharing an $a$ bond and 
Taking $\bf r$ as a unit cell coordinate,  the Majorana fermions can be combined into two \textit{complex} species: bond fermions
$\hat{\chi}^{\dagger}_{\left\langle i j \right\rangle_a}=\frac{1}{2} (\hat{b}_i^a -i \hat{b}_j^a)$ and matter fermions
$\hat{f}_{{\mathbf r}}  =  \frac{1}{2} (\hat{c}_{A {\mathbf r}}+i\hat{c}_{B {\mathbf r}})$ 
%which obey standard anti-commutation relations 
\cite{Baskaran2007}. 
%Here $i,j$ label the neighbouring sites sharing an $a$ bond, and $\mathbf{r}$ is the unit cell coordinate. 
Defining bond operators $\hat{u}_{\langle ij\rangle_a}=i\hat{b}^{a}_{i}\hat{b}^{a}_j$, which commute with $\hat{H}$, the model in terms of 
gauge degrees of freedom and Majorana fermions is
%02
\begin{equation}\label{majorana}
\hat{H}=i \sum_{a,\langle ij\rangle_{a}}J_a\hat{u}_{\langle ij\rangle_{a}}\hat{c}_{i}\hat{c}_{j}.
\end{equation}
The Hamiltonian
$\hat{H}$ has the 
%form of a 
Bogoliubov de-Gennes form
%Hamiltonian
when expressed in terms of %complex 
matter fermion operators $\hat{f}^\dagger_{{\mathbf r}}$ and $\hat{f}_{{\mathbf r}}$
for eigenstates of the gauge fermion operators $\hat{u}_{\langle ij\rangle}$.
%for any given eigenstate of the gauge fermion operators $\hat{u}_{<ij>}$, 
It therefore conserves fermion {\it parity}, but not fermion {\it number}. These features differentiate
our spin dynamics problem from the conventional X-ray edge problems and turn out to be central to our findings. 

The Hilbert space of the Hamiltonian in Eq.~($\ref{majorana}$) can now be decomposed into gauge $|F\rangle$ and  matter $|M\rangle$ sectors, and we denote the ground state of $\hat{H}$ by $|0\rangle=|F_0\rangle\otimes|M_0\rangle$, in which $\hat{u}_{\langle ij\rangle_a}|F_0\rangle=+1|F_0\rangle$ for all bonds and $|M_0\rangle$ is the corresponding ground state of the Majorana hopping problem \cite{Kitaev2006}, whose Hamiltonian $\hat{H}_0$ is obtained from $\hat{H}$ by substituting all $\hat{u}_{\langle ij\rangle_a}$ with their ground-state eigenvalues $+1$.

{\it The dynamical structure factor.} Our objective is to calculate the spin correlation function $S^{ab}_{ij}(t)=  \langle 0| \hat{\sigma}_{i}^{a}(t) \hat{\sigma}_{j}^b(0)|0\rangle$ and 
its Fourier transform in space and time, 
the
%corrseponding
dynamical structure factor
$S^{ab}(\mathbf{q},\omega)$.
%ts Fourier transform in space and time.
%03
%\begin{equation}\label{sfactor}
%S^{ab}(\mathbf{q},\omega)= \frac{1}{N}\sum_{i,j}^{N} e^{-i {\mathbf q} (\mathbf{R}_i-\mathbf{R}_j) } \int_{-\infty}^{\infty} d t e^{i\omega t} S^{ab}_{ij}(t),
%\end{equation}
%where the summation is over $N$ lattice sites $\mathbf{R}_i$. Note that for $\omega>0$ 
%The structure factor 
The latter is
%directly 
proportional to the cross section obtained in an inelastic neutron scattering (INS) experiment, and at $\mathbf{q}=0$ to the signal obtained in electron spin resonance (ESR). 
%The structure factor

The measurement process creates a spin flip, which introduces
a pair of fluxes in adjacent plaquettes (as illustrated in Fig.~\ref{lattice}) and
initiates the dynamical rearrangement of matter fermions in the modified gauge field. 
Because fluxes are static, site off-diagonal spin correlations vanish except for $a$-components of
a nearest-neighbour pair $\langle ij \rangle_a$ \cite{Baskaran2007}. We indicate this using the symbol $\delta_{\langle ij \rangle, a}$. (In the rest of the 
paper we show expressions for the nearest neighbour correlator; the ones for the 
site-diagonal terms are similar.)
Crucially, 
the non-zero contributions to the structure factor can be expressed purely in terms of matter fermions in the ground state flux sector,
subject to a perturbation 
%at $\mathbf{r}=0$ of 
$ \hat{V}_a=-2 i J_a \hat{c}_{i} \hat{c}_{j}$, using the expression
\cite{Baskaran2007}
%03
\begin{equation}
\label{ruegg}
S^{ab}_{ij}(t)  =  - i \langle M_0| e^{i \hat{H}_0 t} \hat{c}_{i}  e^{-i (\hat{H}_0 +\hat{V}_a)t} \hat{c}_{j}|M_0\rangle \delta_{ab} \delta_{\langle ij \rangle, a}~.%~,
\end{equation}
%subject to a perturbation at $\mathbf{r}=0$ of $ \hat{V}_a=-2 i J_a \hat{c}_{i} \hat{c}_{j}$: 
The Hamiltonians $\hat{H}_a=\hat{H}_0 + \hat{V}_a$ and $\hat{H}_0$ differ only in the sign of the Majorana hopping on the $a$-bond, representing insertion of the flux pair. The Lehmann representation of Eq.~(\ref{ruegg}) can be written in the basis of many-body eigenstates of the Hamiltonian $\hat{H}_a$, denoted by $|\lambda\rangle$ with the corresponding energies $E_\lambda$, taking $E_0$ as the ground state energy of $\hat{H}_0$. We choose to work in the fixed gauge in which the eigenvalues of $\hat{u}_{\langle kl\rangle}$ are +1 for all bonds except the one linking the pair of sites $i$ and $j$ that appear in the correlator $S_{ij}^{ab}(t)$.  Then
%04
\begin{multline}
\label{Lehmann}
%\frac
S^{ab}_{ij}(\omega)
=-i\sum_\lambda \langle M_0 |\hat{c}_i | \lambda \rangle \langle \lambda | \hat{c}_j | M_0 \rangle \\ 
\times\delta \left[ \omega - \left( E_\lambda - E_0 \right)\right] \delta_{\langle ij \rangle, a} \delta_{ab}.
\end{multline}

%corresponding to the insertion of a flux pair in the plaquettes sharing this bond (Fig. \ref{lattice}). Thus, $\delta_{\langle ij \rangle, a}$ denotes the non-vanishing of correlations for neighbouring sites only, under the condition that the bond joining them is of type a (Fig. \ref{lattice}). 

\begin{figure}[t]
\includegraphics[width=0.9\columnwidth]{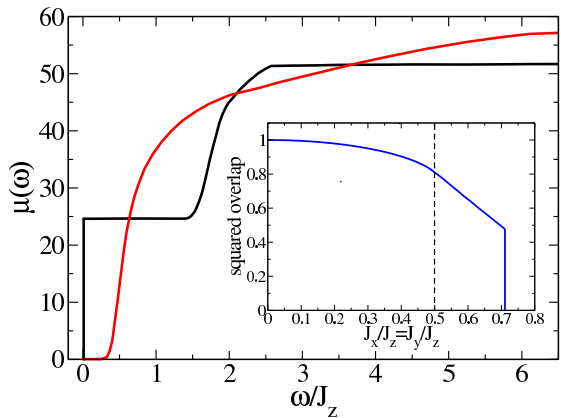}
\caption{%(Colour online). 
{\bf Dynamical transition in inelastic response.}
The integrated structure factor $\mu(\omega)=\int_0^{\omega} d\omega' \sum_a S^{aa}(%{\bf q}=
0,\omega')$ in the gapless $J_x=J_y=J_z$ (red), and the gapped $J_x/J_z=J_y/J_z=0.15$ (black) phases.
% In the latter case the step around $\omega=2 J_z$ arises due to contributions of the weak bonds ($a{=}x,y$). 
The sharp step in the black curve at low $\omega$ at the two-flux gap $\Delta \simeq J_x^4/8J_z^3$ is due to the $\delta$-function contribution of the strong bond correlator ($a=z$).
Inset: The dependence of the ground state overlap $|\langle \lambda_0|E_0\rangle|^2$,
proportional to the $\delta$-function contribution, on interaction strength,
along the line $J_x/J_z = J_y/J_z$.
 %(rescaled by $1/8 \pi$ ) calculated from finite size numerics 
 %for a fixed value of $J_z=1$
}\label{Parity}
\end{figure}

{\it Results.} We start our discussion of the results displayed in Figs.~\ref{threepanels} and \ref{Parity} by explaining the salient qualitative features in terms of the selection rules imposed by the fractionalisation of the electrons into fluxes and Majoranas. 
%Recall, the vanishing of correlations beyond the nearest-neighbour-distance follows from the fact that a spin flip (or, more precisely, the action of a spin operator) involves the creation of a flux pair (Fig. \ref{lattice}), which needs to be removed by the action of the second spin operator \cite{Baskaran2007}. This is implicitly taken into account by restricting the intermediate states $| \lambda \rangle$ in (\ref{Lehmann}) to those within the appropriate flux sector.
It is instructive to do so using Eq.~(\ref{Lehmann}), from which the central aspects of the response can be read off, and we relegate an explanation of the numerically exact solution of Eq.~(\ref{ruegg}) to the supplementary material.

First, in both gapped and gapless phases, response vanishes below the two-flux gap $\Delta=E_{\lambda_0}-E_0$, 
%which is 
the difference between the ground state energies in a system with and without the flux pair. (At the isotropic point $\Delta\simeq 0.26J$ \cite{Kitaev2006}.) It is remarkable that in an INS experiment the response of a gapless QSL will show an excitation gap which is directly related to the emergent gauge field.

Above the gap $\Delta$, the response thus reflects the physics of the matter 
%(Majorana) 
sector, and our analysis uncovers an entirely new structure in the phase diagram of the Kitaev model, Fig.~\ref{lattice}b. An important consequence of the fact that $\hat{H}$ in Eq.~(2) conserves matter fermion parity is that the non-zero contributions to Eq.~(\ref{Lehmann}) come only from excited states $|\lambda \rangle$ with parity opposite to the ground state $|M_0\rangle$. As a result, two distinctively different alternatives arise: either (I) the ground states of $\hat{H}_0$ and $\hat{H}_a$ have the same parity, in which case the states $|\lambda \rangle$ must contain an odd number of excitations, or (II) the ground states have opposite parity and $|\lambda \rangle$ contains an even number of excitations, a condition that is also fulfilled by the ground state of $H_a$.

For (I) [Fig.~\ref{threepanels} a-d], single particle excitations dominate the response, which is broad in energy, so that its amplitude is appreciable only within the matter fermion bandwidth. Indeed, only about $2.5 \%$ of the signal at the symmetric point arises from multi-particle contributions (see supplementary material) in stark contrast to the case of the Heisenberg chain \cite{ruegg}, where the corresponding number is almost $30 \%$.

For (II) [Figs.~\ref{threepanels} e and f], in striking opposition, the response includes a finite-weight $\delta$-function component in $\omega$ at the difference $\Delta$ in ground state energies, since the corresponding matrix element is finite. It is a remarkable and unexpected finding that -- despite fractionalisation -- the INS response has a component sharp in energy (displayed in Fig.~\ref{Parity} b).
Note that the location in the phase diagram of the dynamical transition at which this sharp response appears is distinct from the ground state phase boundary: it lies entirely within the gapless phase (Fig. \ref{lattice} b).

{\it Discussion.} 
%{\it Connection to quantum quench:} 
Formally, Eq.~(\ref{ruegg}) represents an example of a quantum quench: it involves the overlap between a state $\langle M_0 |\hat{c}_i$ that is
simple in terms of  $\hat{H}_0$ (a superposition of single-particle excitations) and
a similar state $\hat{c}_{j}|M_0\rangle$ after the latter has evolved for time $t$ under a different Hamiltonian $\hat{H}_a$.
%The ground state of $\hat{H}_0$, from time $\sigma$ onwards, evolves according to $\hat{H}_a$. 
The broad features of the resulting response of the Majorana fermions above $\Delta$ are a result of this quench. Quite surprisingly, this can be well approximated by replacing the instantaneous flip of the bond by an adiabatic, rather than sudden, switching-on of the potential $\hat{V}_a$. This amounts to replacing $| M_0 \rangle$ in Eq.~(\ref{Lehmann}) by the Majorana ground state {\it in the presence of the fluxes}. One can show that in the limit of low energies, the matter fermion eigenstates are, in fact, insensitive to the flux addition, so that the resulting approximation (dashed line, Fig. \ref{threepanels} b) becomes exact as $\omega$ approaches $\Delta$. 

It is interesting to compare the energy dependence of the structure factor with the density of states for matter fermions (Fig.~\ref{threepanels} b and d). Response is substantial over the entire single-particle band width (shifted in energy by $\Delta$), with linear onset above the gap. 
However, as a qualitative signature of the effect of gauge fluxes on
matter fermion dynamics, the response is far from being simply proportional to the density of states.  
Instead, the peak in the latter at $2J_z$ due to the van Hove singularity [see inset to Fig.~\ref{threepanels} b] yields {\it a dip} in the response. Away from the symmetric point there are 
%the saddle points of the single-particle dispersion become inequivalent, generating 
two van Hove singularities in the density of states, and in addition there is a distinct response
for differently orientated spin pairs,
%along $z$ and $x,y$ bonds is different 
showing one or two minima in the corresponding dynamical susceptibility (Fig. \ref{threepanels} d).

Despite the formal similarities between the time-dependent correlator Eq.~(\ref{ruegg}) and the X-ray edge problem, the physics arising from it is quite different. First, depending on the exchange $J_a$ one can study a local quantum quench in either gapless or gapped phases, the latter not presenting the possibility of low-energy fermionic excitations. Second, for inequivalent values of $J_a$ the correlators for different spin components are different. Third, the Majorana fermions in our calculation arise due to fractionalisation of  spin degrees of freedom as emergent particles. Fourth,  they have not number, but only parity conservation, and their dispersion exhibits Dirac cones. Finally, the $\delta$-function response (Fig. \ref{Parity} inset) is diametrically opposed to Anderson's orthogonality catastrophe, which would correspond to a vanishing signal at $\Delta$. 

%{\it Experiment:} 

The Kitaev model is a representative of a class of $\mathbb{Z}_2$ spin liquids coupled to (gapped or gapless) 
Majorana fermions. The qualitative features described above should therefore be characteristic of this broad class of topological
states. While a potential cold atom realisation will likely 
%going to 
harbour few perturbations to the Kitaev Hamiltonian (\ref{HkitaevSpin}),
magnetic materials 
%would will most likely 
usually include other terms, as extensively discussed for 
Kitaev-Heisenberg models following Ref.~\onlinecite{Jackeli2009}. Both the flux gap and the fermion parity underpinning our results
are robust to such perturbations. Just as in the analogous case of the Heisenberg chain 
%in $d=1$
 \cite{ruegg}, where integrability %in reality
is %also 
imperfect in reality
%not perfect 
but all qualitative features are well-observed experimentally, so we similarly expect quantitative changes such as 
a small degree of smearing out of the $\delta$-function response or a more gradual onset
of the signal around $\Delta$. 
Crucially, the central features we have discovered will be visible as fingerprints betraying the presence of fractionalised Majorana fermions and emergent gauge fluxes in INS and ESR experiments. 

%\begin{figure}
%\begin{centering}
%\includegraphics[width=1.0\columnwidth]{Fig4new.eps}
%\par\end{centering}
%\caption{Correlation function $S_{AB}^{zz}(t)$ is a product of connected Greens function $G_c$ and the closed loop contribution $L(t)$, see Eq.~(\ref{CorrFctFermiEdge}). The time-dependence of two connected Green functions is shown at the isotropic point $J_{x,y,z}=1$ in the gapless phase, and $|L(t)|$ saturates to a finite value at long times (see inset).\label{realtime}}
%\end{figure}

\vspace{1cm}

\noindent\textbf{\textsf{Acknowledgements}}\\

We thank F.~H.~L.~Essler and G.~Baskaran for helpful discussions. J.K.~acknowledges support from the Studienstiftung des deutschen Volkes, the IMPRS Dynamical Processes in Atoms, Molecules and Solids, DFG within GRK1621. J.T.C.~is supported in part by EPSRC Grant No. EP/I032487/1. D.K.~is supported by EPSRC Grant Nos.~EP/J017639/1 and EP/J009636/1, and acknowledges the GGI workshop ``New quantum states of matter in and out of equilibrium'', where part of this work was done. This collaboration was supported by the Helmholtz Virtual Institute ``New States of Matter and their Excitations''.\\

%\vspace{1cm}

%\vspace{1cm}

% \noindent\textbf{\textsf{Author contributions}}\\
% All authors contributed equally to the work.\\
% 
% 
% \noindent\textbf{\textsf{Additional information}}\\
% The authors declare no competing financial interests. Correspondence and requests for materials should be addressed to J.K.

%\appendix
%\section{Appendix}      
%\begin{itemize}
%\item overlap discussion and presence of bound state depending on bond direction; connection to two 'origins' of orthogonality mentioned in Anderson's paper
%\item asymptotic behaviour of Green functions etc.; justification for validity of quasi-adiabatic heuristic
%\item broader discussion of difference to Fermi edge problem
%\item connections to other work (babanov?)
%\end{itemize}
%
%Other things which can go in, or be treated later, space and result availability permitting: 
%\begin{itemize}
%\item overlap from triple integral 
%\item application to graphene
%\item connection to hopping problem (failure to connect different energy sectors)
%\item flux energy from scattering calculation
%\item 4,6,8 $J_z$ numerics in the gapped phase
%\item long-time behaviour of response functions
%\item general rant about Kitaev's dynamics collaborators
%\end{itemize}

%\newpage
\vspace{1cm}
\appendix

\noindent\textbf{\textsf{Supplementary material}}\\

\noindent In this supplementary material we present an outline of our numerically exact evaluation of the spin-correlators in Eq.~(\ref{ruegg}), and of the calculation of the few-particle response from the Lehmann representation given in Eq.~(\ref{Lehmann}).

\noindent\textit{\textbf{Definitions.}} The foundation
for Kitaev's solution of the model \cite{Kitaev2006} is the representation of spins using Majorana fermions in an enlarged Hilbert space. Calculation of observables requires projection
onto the physical subspace. Since this step is standard \cite{Baskaran2007}, we do not discuss it here. As described in the main text,
we proceed by combining the Majorana operators into two independent complex fermion species
\begin{equation}
\hat{\chi}^{\dagger}_{\left\langle i j \right\rangle_a}=\frac{1}{2} (\hat{b}_i^a -i \hat{b}_j^a),\ \ 
\hat{f}_{{\mathbf r}}  =  \frac{1}{2} (\hat{c}_{A {\mathbf r}}+i\hat{c}_{B {\mathbf r}}),
\end{equation}
termed bond and matter fermions, respectively.
%These obey standard fermionic anticommutation relations
%$\{\hat{\chi}_{\alpha},\hat{\chi}^{\dagger}_{\beta}\}=\delta_{\alpha\beta},\ \{\hat{f}_{\mathbf{r}},\hat{f}^{\dagger}_{\mathbf{r}'}\}=\delta_{\mathbf{r}\mathbf{r}'}.$
The bond operators $\hat{u}_{\langle ij\rangle_a}=i\hat{b}^{a}_{i}\hat{b}^{a}_j$ can  then be expressed %terms of bond fermions $\hat{\chi}^{\dagger}_{\left\langle i j \right\rangle_a}$ 
as
%\begin{equation} 
$\hat{u}_{\langle ij\rangle_a}=2\hat{\chi}_{\left\langle i j \right\rangle_a}^{\dagger}\hat{\chi}_{\left\langle i j \right\rangle_a}-1.$
%\end{equation}
Using this representation we can diagonalise the Hamiltonian of Eq.~($\ref{majorana}$), on a torus with $N$ unit cells, by a Fourier transform $\hat{f}_{{\mathbf r}}=\frac{1}{\sqrt{N}}\sum_{\mathbf q} e^{i{\mathbf q}{\mathbf r}}\hat{f}_{\mathbf q}$, followed by a Bogoliubov transformation $\hat{f}_{\mathbf q}=\cos\theta_{\mathbf q}\hat{a}_{\mathbf q}+ i\sin\theta_{\mathbf q}\hat{a}^{\dagger}_{-\mathbf q}$. This gives the Hamiltonian in the flux-free matter sector as
\begin{eqnarray}
\label{HamiltonianDiag}
\hat{H}_0=\sum_{\mathbf {q}\in \mathrm{BZ}} |s_{\mathbf q}|(2 \hat{a}_{\mathbf q}^{\dagger}\hat{a}_{\mathbf q}-1),
\end{eqnarray}
where $s_{\mathbf q}=J_z+J_x e^{i {\mathbf q} {\mathbf{n}_1}}+J_y e^{i {\mathbf q} {\mathbf{n}_2}}$. The lattice vectors $\mathbf{n}_1,\mathbf{n}_2$ are defined in Fig. $\ref{lattice}$ a, and $\tan{2 \theta_{\mathbf q}}=-\mathrm{Im}[s_\mathbf{q}]/\mathrm{Re}[s_\mathbf{q}]$. The ground state of the matter sector $|M_0\rangle$ is defined by the condition that $%\hat{a}_\mathbf{q}^{\dagger}
\hat{a}_{\mathbf{q}}|M_0\rangle=0$ for all $\mathbf{q}$ in the Brillouin zone, which gives the ground state energy $E_0=-\sum_{\mathbf{q}}|s_{\mathbf{q}}|$ \cite{Kitaev2006}.

In the following we need a number of different Green functions (GF), which we list here for convenience.
%, with definitions postponed. 
We denote the time-ordered GF  \cite{AGD} for $\hat{H}_0$ by $g_0(t)$, its Fourier transform (FT) by $g_0(\omega)$, and the corresponding advanced GF by $g^{a}_0(\omega)$. 
For the quantum quench problem the time-ordered GF $g(t,0)$ and 
the transient GF $g(\tau_1,\tau_2;t)$ appear, as well as the connected part $g_c(\tau_1,\tau_2;t)$ and the FT $g_c(\omega;t)$ of this. Finally, we use the adiabatic approximation $\tilde{g}^a_0(\omega)$
to the latter.

%\noindent\textbf{\textit{Green-functions.}} In the following we need a number of different Green-functions (GF) which we list here, while postponing their discussion to a later stage. First, the standard single-particle time-ordered GF \cite{AGD} in the interaction representation with the free Hamiltonian $\hat{H_0}$ is defined as
%\begin{equation}
%g_0(t)=-i\langle M_0|T\{\hat{f}_0(t)\hat{f}_0^{\dagger}(0)\}|M_0\rangle,
%\end{equation}
%where $T\{\ldots\}$ denotes the time-ordering. 
%
%After transforming to the frequency representation such that $g_0(\omega)=\int g_0(t)e^{i\omega t} dt$, with the infinitesimal $\delta=+0$, we obtain
%\begin{equation}\label{spGF}
%g_0(\omega)=\frac{1}{N}\sum_{\mathbf{q}\in\mathrm{BZ}}\left[\frac{\cos^2{\theta_{\mathbf{q}}}}{\omega-2|s_\mathbf{q}|+i\delta}+\frac{\sin^2{\theta_{\mathbf{q}}}}{\omega+2|s_\mathbf{q}|-i\delta}\right].
%\end{equation}

%We denote an advanced GF by index $a$ as $g^{a}_0(\omega)$, whose definition in terms of the time-ordered GF is given in a standard way $\mathrm{Re}[g^{a}_0(\omega)]=\mathrm{Re}[g_0(\omega)]$ and $\mathrm{Im}[g^{a}_0(\omega)]=-\mathrm{Im}[g_0(\omega)]\ \mathrm{sign}(\omega)$. We also use an adiabatic GF $\tilde{g}^{a}(\omega)$, a transient GF $g(\tau_1,\tau_2;t)$ together with a connected part $g_c(\tau_1,\tau_2;t)$ and its Fourier transform $g_c(\omega;t)$.

\noindent\textit{\textbf{Exact approach.}} Our main task is to evaluate %the expression in 
Eq.~(\ref{ruegg}).
% numerically exactly. In doing so 
For this we use a mapping onto the X-ray edge problem \cite{Baskaran2007,Tikhonov2010,Tikhonov2011}, which gives a connection to a very well-studied area in many-body physics and allows one to obtain an expression for the correlators that is suitable for high-precision numerical evaluation. 

Consider for definiteness the correlator on the $z$-bond $S^{zz}_{A0B0}(t)=-i\langle M_0|\hat{c}_{A0} e^{-i(\hat{H}_0+\hat{V}_z)t}\hat{c}_{B0}|M_0\rangle$, where the explicit form of the potential is $\hat{V}_z=-2iJ_z\hat{c}_{A0}\hat{c}_{B0}$. We express the correlators entirely in terms of $\hat{f}$-fermions and it is convenient to work in the interaction representation, where the time dependence of the operators is given by the Hamiltonian $\hat{H}_0$ so that $\hat{f}_{\mathbf{r}}(t)=e^{i\hat{H}_0 t}\hat{f}_{\mathbf{r}}e^{-i\hat{H}_0 t}$. Then $\hat{V}_z$ assumes the {\it local} form
\begin{equation}\label{VZ}
\hat{V}_z(t)=v\left[\hat{f}_0^{\dagger}(t)\hat{f}_0(t)-\frac{1}{2}\right],
\end{equation}
with $v=-4J_z$. The potential in Eq.~(\ref{VZ}) is similar to the one that appears in the context of the X-ray edge problem (see e.g.~\cite{Gogolin2004}). Its essentially point-like form makes an exact solution possible.  Now we can use the fermion representation to express spin correlators in terms of %standard \cite{AGD} 
time-ordered GFs,
\begin{equation}
g(t,0)=-i\langle M_0|T\{\hat{f}_{0}(t)\hat{f}^{\dagger}_{0}(0)\hat{S}(t)\} |M_0\rangle,\label{GFp1}
%g(0,t)&=&-i\langle M_0|T\{\hat{f}_{0}(0)\hat{f}^{\dagger}_{0}(t)\hat{S}(t)\} |M_0\rangle,\label{GFp2}
\end{equation}
with the scattering matrix $\hat{S}(t)$ defined as
\begin{equation}
\hat{S}(t)=T\exp{\left[-i\int_{0}^{t}\hat{V}_z(\tau)d\tau\right]}.
\end{equation}
The GF $g_0(t)$ is also given by Eq.~(\ref{GFp1}), but with $\hat{S}(t)=1$.
%In terms of these GFs, 
The spin-correlator then reads
\begin{equation}
S^{zz}_{A0B0}(t)=i[g(t,0)+g(0,t)].
\end{equation} 

%The ground-state $|M_0\rangle$ is a zero particle number state of $\hat{a}_\mathbf{q}$ operators, and 
Because $\hat{f}$-fermions are given as a linear superposition of $\hat{a},\hat{a}^{\dagger}$, one would in general expect to have contributions from anomalous correlators $\langle {M}_0| \hat{f}_0^{\dagger}(t_1) \hat{f}^{\dagger}_0(t_2)|{M}_0\rangle$. What makes the $\hat{f}$-representation useful 
%in the case of the Kitaev honeycomb model 
is the simplifying feature that these anomalous averages vanish for any $t_1,t_2 $, as a consequence of inversion symmetry and gauge invariance. 
% This can be most straightforwardly seen in the Majorana representation and using the property of $\hat{c}_{A/B}$ operators, that they generate unitary transformations of the Hamiltonian $\hat{H}$ flipping the signs of three bonds adjacent to $A/B$ site. The bond configurations generated by $\hat{c}_A$ and $\hat{c}_B$ are related by a gauge transformation and correspond to the same flux configuration.

The task of calculating the GFs in Eq.~(\ref{GFp1}) is thus \textit{equivalent} to the 
%standard 
X-ray edge problem in which a local potential is switched on and off at times $0$ and $t$. The problem can therefore be formulated in terms of an integral equation as is done in the X-ray edge case \cite{Gogolin2004}. The bare GF $g_0(t)$ in the Kitaev model is, however, non-standard, %Eq.~(\ref{spGF}), 
reflecting the presence of Dirac points or a gap, rather than a Fermi surface \cite{Tikhonov2010,Tikhonov2011}.

Consider, for example, $g(t,0)$. Using Wick's theorem we can write this as a product
$g(t,0)=g_c(t,0)\mathcal{L}(t)$
of the connected GF $g_c(t,0)$ and the loop contribution $\mathcal{L}(t)=\langle M_0|\hat{S}(t)|M_0\rangle$. The latter is equal to the overlap between the state without fluxes $\langle M_0|$ and a state $\hat{S}(t)|M_0\rangle$ after evolution to time $t$ with fluxes (potential $\hat{V}_z$) introduced at time $0$. The loop contribution can be expressed purely in terms of connected GFs \cite{Gogolin2004} and we concentrate on finding the latter using an integral equation for $g_c(t,0)$. This takes the form of the equation
\begin{equation}\label{inteq}
g_c(t,0)=g_0(t)+v\int_{0}^{t}g_0(t-\tau)g_c(\tau,0)d\tau\,,
\end{equation}
which was solved exactly in the long-time limit for a Fermi-liquid with $g_0(t)\sim 1/t$ in a seminal paper by Nozieres and De Dominicis, Ref.~\cite{Nozieres1969}. 

In the Kitaev honeycomb model the asymptotics of $g_0(t)$ are not of the standard $1/t$ form. Since that special form is crucial to the exact solution of Eq.~(\ref{inteq}), we cannot use the same approach. Instead, we use methods developed in the theory of singular integral equations \cite{mush,greb} to recast Eq. (\ref{inteq}) into a version suitable for numerical evaluation, as described below.

Before embarking on this, we 
%make a couple of side remarks. First, note that in the limit $t\to\infty$ Eq.~(\ref{inteq}) is of the Wiener-Hopf type, which allows one to find the overlap between the states with and without fluxes analytically along the lines of the calculation presented in Chapter 26.V of Ref.~\cite{Gogolin2004}. Note also that if we suppose 
remark that if the two-flux potential is switched on and off adiabatically, instead of suddenly at times $0$ and $t$, the corresponding integral equation obtained from Eq.~(\ref{inteq}) by taking the integration limits $(-\infty,\infty)$ is readily solved via FT. This leads to the result for the adiabatic GF
\begin{equation}
\tilde{g}^{a}_0(\omega)=\frac{g^{a}_0(\omega)}{1-v g^{a}_0(\omega)}.
\label{AdiaGF}
\end{equation}
Here, the advanced GF $g^{a}_0(\omega)$ is related to the time-ordered one by $\mathrm{Re}[g^{a}_0(\omega)]=\mathrm{Re}[g_0(\omega)]$ and $\mathrm{Im}[g^{a}_0(\omega)]=-\mathrm{Im}[g_0(\omega)]\ \mathrm{sign}(\omega)$.
The response obtained using the adiabatic GF (\ref{AdiaGF}) can be compared with the one from the exact calculation (see Fig.~\ref{threepanels} b). One can show that in the Kitaev model this adiabatic approximation becomes exact in the low energy limit due to the vanishing density of states of matter fermions. In addition, Eq.~(\ref{AdiaGF}) encodes the fact that the peak of the van Hove singularity in the density of states yields a dip in the response.
%, see Fig.~\ref{threepanels}b.

\noindent\textit{\textbf{Solution of the integral equation.}} In order to calculate the structure factor as a function of frequency we need the solution to Eq.~(\ref{inteq}) in the whole time domain. 
The sharp integration boundaries and large integration range for large $t$ mean that
this equation is not convenient for numerical evaluation. 
%In addition it is also hampered by the two singularities due to the sharp integration boundaries $0$ and $t$. 
Instead, we transform it
%Eq.~(\ref{inteq}) 
into the frequency domain with the advantage of having \textit{finite} integrals due to finite bandwidth of the single-particle density of states.% $\sim\mathrm{Im}[g^{a}_0(\omega)]$. 

The basic steps follow closely the approach of Ref.~\cite{greb}.
%, which was developed in the context of the X-ray edge problem. 
The FT of Eq.~(\ref{inteq}) to the frequency domain is taken with respect to the time argument of the GFs, while keeping the time $t$ in the integral fixed, i.e.~$g_c(\omega;t)=\int g_c(\tau,0;t)e^{i\omega \tau}d\tau$. Since the adiabatic approximation provides a good description at small energies, it is natural to write the GFs as a product $g_c(\omega;t)={g}_0(\omega)\varphi_{\omega}(t)$, which gives the equation
\begin{equation}\label{IG}
\varphi_\omega(t)=1+v\int_{-\infty}^{+\infty}\frac{d\omega'}{2\pi i}\frac{e^{i(\omega-\omega')t}-1}{\omega-\omega'}\varphi_{\omega'}(t).
\end{equation}
Next, by using analytic properties of GFs it is possible to write the r.h.s of Eq.~(\ref{IG}) in terms of integrals with retarded and advanced contributions, and finally bring it to the form where the kernel of the equation is proportional to the density of states, and in particular only has a finite bandwidth. 

However, in this form the kernel has a $1/\omega$ singularity  and requires further simplifications.
As explained in great detail in Ref.~\cite{mush}, such a singularity can be removed by acting on the equation with a specially constructed integral operator. After performing this procedure, and introducing two auxiliary functions
\begin{eqnarray}
f_\omega(t)&=&1-\frac{v}{\pi}\int_{-\infty}^{\infty}d\omega'\ \frac{e^{i(\omega-\omega')t}-1}{\omega-\omega'}\mathrm{Im}[\tilde{g}^{a}_0(\omega')],\\
R_{\omega\omega'}(t)&=&\frac{f(\omega)-e^{i(\omega-\omega')t}f(\omega')}{\omega-\omega'},
\end{eqnarray}
we can finally express the Eq.~(\ref{IG}) in the form
\begin{equation}\label{IGW}
\varphi_{\omega}(t)=f_\omega(t)-\frac{v}{\pi}\int_{-\infty}^{0}d\omega'\ R_{\omega\omega'}(t)\mathrm{Im}[g^{a}_0(\omega')]\varphi_{\omega'}(t),
\end{equation}
and from the solution of this equation we obtain the connected GF in the time-domain
\begin{equation}
g_c(t,0)=-\frac{i}{\pi}\int^{\infty}_{0}d\omega\ e^{-i\omega t}\varphi_{\omega}(t)\mathrm{Im}[g^{a}_0(\omega)],
\end{equation}
as well as the corresponding loop contribution
\begin{equation}
\mathcal{L}(t)=\exp\left[-\frac{iv}{\pi}\int_0^t d\tau\int_{-\infty}^{0}d\omega\ \varphi_{\omega}(\tau)\mathrm{Im}[g_0^{a}(\omega)]\right].
\end{equation}
In practice we discretise the Eq.~(\ref{IGW}) in the frequency domain and solve the resulting system of equations numerically. We find that with 8001 discretisation steps, errors are not visible on the scale
of our figures.
%The error is controlled by the discretisation step for a given system size, which can be taken to infinity for a thermodynamic limit.

\noindent\textbf{\textit{Few-particle response.}} 
While the treatment summarised above gives a practical route to computing the full response, it is also 
interesting to analyse what proportion of this is contributed by states with small numbers of excitations.
To do so, note that
% the potential $\hat{V}_z$ is simply local, which gives us a way of calculating the exact response. 
the ground state $|M_0\rangle$ is a vacuum of $\hat{a}$-fermions: in order to calculate the overlap and the few-particle response it is convenient to express all other operators in the same representation. In this basis the potential acquires anomalous terms $\sim\hat{a}^{\dagger}\hat{a}^\dagger$. 

The Hamiltonian $\hat{H}_0+\hat{V_a}$ is quadratic and can be diagonalised by a Bogoliubov transformation \cite{ripka} to new fermion operators
\begin{equation}
\hat{b}_{\lambda}=\sum_{\mathbf{q}}X^*_{\lambda \mathbf{q}} \hat{a}_{\mathbf{q}}+Y^*_{\lambda \mathbf{q}}\hat{a}_{\mathbf{q}}^{\dagger},
\end{equation}
 where $X,Y$ are the transformation matrices. The single-particle eigenstates are given by $|\lambda\rangle=\hat{b}^{\dagger}_{\lambda}|\lambda_0\rangle$, where the ground state of the two-flux sector $|\lambda_0\rangle$ can be written in terms of $\hat{a}$-fermions and the ground state $|M_0\rangle$ in the matter sector at zero flux  as
\begin{equation}
|\lambda_0 \rangle = [X^{\dagger}X]^{\frac{1}{4}} e^{-\frac{1}{2}\hat{\mathbf{a}}^{\dagger}X^{* -1} Y^* \hat{\mathbf{a}}^{\dagger}}|M_0\rangle.
\end{equation} 
The multiply-excited states can similarly be obtained as $\hat{b}^{\dagger}_{\lambda_{1}}\hat{b}^{\dagger}_{\lambda_{2}}\ldots|\lambda_0 \rangle$. Using these definitions
we can immediately compute the overlap $|\langle\lambda_0|M_0\rangle|=\sqrt{|\mathrm{det}X|}$, and evaluate the single-particle contribution to the response functions, e.g.~for the combined correlator $\sum_{ij=A0,B0}S^{zz}_{ij}(\mathbf{q}=0,\omega)$  we have
\begin{multline}
S^{zz}_{[1]}(\mathbf{q}=0,\omega)=\frac{2}{N}|\det{X}|\sum_{\lambda, \mathbf{qq'}}\delta[\omega-(E_\lambda^{1}-E_0)]\\
\times\cos{\theta_{\mathbf{q}}}X^{-1}_{\mathbf{q}\lambda}[X^{-1}]^{\dagger}_{\lambda\mathbf{q'}}\cos{\theta_{\mathbf{q'}}},\end{multline}
where the sum on $\lambda$ runs over single-particle energies $E_\lambda^1$ in the system with inserted fluxes. Comparison of the results of this calculation with those of Fig.~\ref{Parity} shows that the single-particle contribution is the dominant one, as asserted in the main text.

\end{document}